\documentclass[9pt,twocolumn,twoside]{pnas-new}

\templatetype{pnasresearcharticle} 

\title{Superdiffusion of quantized vortices uncovering scaling behavior of quantum turbulence}

\author[a,b]{Yuan Tang}
\author[a,b]{Shiran Bao}
\author[a,b,*]{Wei Guo}

\affil[a]{National High Magnetic Field Laboratory, 1800 East Paul Dirac Drive, Tallahassee, Florida 32310, USA}
\affil[b]{Mechanical Engineering Department, Florida State University, Tallahassee, Florida 32310, USA}

\leadauthor{Tang}

\authorcontributions{ W.G. designed the research; Y.T. performed the research; Y.T., S.R.B., and W.G. analyzed the data and wrote the paper.}
\authordeclaration{The authors declare no competing interest.}
\correspondingauthor{\textsuperscript{*}To whom correspondence should be addressed. E-mail: wguo@magnet.fsu.edu}

\keywords{quantum turbulence $|$ superfluid $|$ superdiffusion $|$ particle tracking velocimetry $|$ scaling laws}

\begin{abstract}
Generic scaling laws, such as the Kolmogorov's 5/3-law, are milestone achievements of turbulence research in classical fluids. For quantum fluids such as atomic Bose-Einstein condensates, superfluid helium, and superfluid neutron stars, turbulence can also exist in the presence of a chaotic tangle of evolving quantized vortex lines. However, due to the lack of suitable experimental tools to directly probe the vortex-tangle motion, so far little is known about possible scaling laws that characterize the velocity correlations and trajectory statistics of the vortices in quantum-fluid turbulence (QT). Acquiring such knowledge could greatly benefit the development of advanced statistical models of QT. Here we report an experiment where a tangle of vortices in superfluid $^4$He are decorated with solidified deuterium tracer particles. Under experimental conditions where these tracers follow the motion of the vortices, we observed an apparent superdiffusion of the vortices. Our analysis shows that this superdiffusion is not due to L\'{e}vy flights, i.e., long-distance hops that are known to be responsible for superdiffusion of random walkers. Instead, a previously unknown power-law scaling of the vortex-velocity temporal correlation is uncovered as the cause. This finding may motivate future research on hidden scaling laws in QT.
\end{abstract}

\begin{document}
	
\maketitle
\thispagestyle{firststyle}
\ifthenelse{\boolean{shortarticle}}{\ifthenelse{\boolean{singlecolumn}}{\abscontentformatted}{\abscontent}}{}

\dropcap{Q}uantum fluids, such as superfluids, superconductors, and Bose-Einstein condensates (BECs), exhibit macroscopic quantum coherence that is responsible for their dissipationless motion \cite{Tilley-book}. In these quantum fluids, all rotational motion is sustained by quantized vortex lines, i.e., line-shaped topological defects characterized by a circulating flow of particles with a discrete circulation $\kappa=h/m$, where $h$ is Planck's constant and $m$ is the mass of the particle.

Turbulence in quantum fluids, i.e., quantum turbulence (QT), can be induced by a tangle of interacting vortex lines \cite{vinen-2002-JLP}. These vortex lines evolve chaotically under their self- and mutually induced velocities and can reconnect when they move across each other \cite{Donnelly-1991-B}. The underlying science of QT is broadly applicable to a variety of coherent physical systems, such as coherent condensed matter systems (e.g., superfluid $^3$He and $^4$He \cite{Barenghi-2001-book}, atomic and polariton BECs \cite{Madeira-ARCMP-2020}, and type-II superconductors \cite{Larbalestier-2001-Nature}), cosmic systems (e.g., neutron-pair superfluid in neutron stars \cite{Greenstein-1970-Nature, Andersson-2007-MNRAS}, cosmic strings in the Abelian-Higgs model \cite{Zurek-1985-Nature}, and possible axion dark matter BECs in galactic halos \cite{Sikivie-2009-PRL}), and even complex light field \cite{Alperin-2019-PRL}. Insight into the generic scaling laws that characterize the evolution of quantized vortex tangles can inform statistical models of QT, which could have a broad significance spanning multiple branches of physics.

QT research has been conducted mostly in superfluid $^3$He and $^4$He due to the material's accessibility and the wide range of length scales involved in their turbulence behaviors \cite{Barenghi-2014-PNAS}. Nevertheless, despite extensive theoretical and numerical studies of the vortex-line dynamics in superfluid helium \cite{Schwarz-1988-PRB,Adachi-2010-PRB,Baggaley-2012-PRL}, past experimental research has largely been limited to the measurements of spatially-averaged quantities such as the vortex-line density $L$ (i.e., length of vortices per unit volume) \cite{Stalp-1999-PRL, Eltsov-2007-PRL, Walmsley-2014-PNAS} or local pressure and temperature variations \cite{Maurer-1998-EPL, Bradley-2011-NP}. Important statistical properties of a fully-developed vortex tangle, such as the vortex-velocity correlations and their trajectory statistics, remain largely unexplored due to the lack of experimental tools for probing the vortex-line motion.

A breakthrough has been made in recent years with the development of quantitative flow visualization techniques \cite{Guo-2014-PNAS}. In particular, by decorating the vortices in superfluid $^4$He (He II) with solidified hydrogen particles, Bewley \emph{et al.} demonstrated direct vortex-line visualization \cite{Bewley-2006-Nature}. Since then, vortex-line reconnections and Kelvin-wave excitations on individual vortices have been filmed \cite{Bewley-2008-PNAS,Fonda-2014-PNAS,Fonda-2019-PNAS}. Nevertheless, visualization data showing the real-time evolution of a complex vortex tangle are still lacking, which impedes the development of reliable statistical models for describing QT \cite{Nemirovskii-2013-PR}.

\begin{figure*}[t]
\centering
\includegraphics[width=1.0\linewidth]{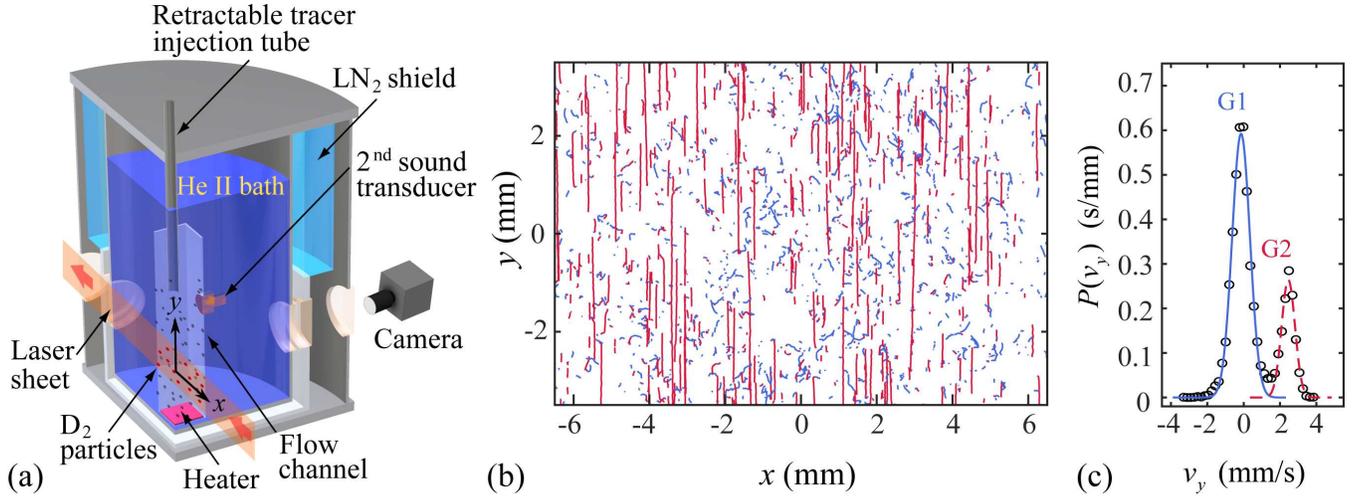}
\caption{(a) A schematic diagram of the experimental setup. (b) Representative trajectories obtained at $T=$1.85 K and $q=$38 mW/cm$^2$ in He II for G1 particles trapped on vortices (blue) and G2 particles entrained by the normal fluid (red). (c) The corresponding streamwise particle-velocity distribution, where the solid and the dashed curves represent Gaussian fits to the data.}
\label{Fig1}
\end{figure*}

In our recent experiment on He II QT driven by an applied heat current, we seeded the fluid with solidified deuterium (D$_2$) tracer particles and observed that a group of particles could remain trapped on the tangled vortices \cite{Mastracci-2017-JLTP,Mastracci-2018-PRF,Mastracci-2019-PRF,Mastracci-2019-PRF-2}. By applying a separation scheme in data analysis \cite{Mastracci-2018-PRF}, we were able to track solely these trapped particles and therefore could directly probe the vortex-tangle dynamics. In this paper, we discuss our study of the apparent diffusion of these trapped particles under experimental conditions where they faithfully follow the motion of the evolving vortices.

We report the first observation of a superdiffusion of the vortices in the tangle when their root mean squared displacement (RMSD) is less than the mean inter-vortex distance $\ell$=$L^{-1/2}$. Surprisingly, our analysis shows that this superdiffusion is not due to L\'{e}vy flights (i.e., randomized, long-distance hops) that are known to be responsible for superdiffusion in various physical and non-physical systems \cite{Zaburdaev-2015-RMP}. Instead, we reveal that a previously unknown power-law scaling of the vortex-velocity temporal correlation is the cause. The derived power-law exponent appears to be temperature and vortex-line density independent, suggesting that the observed scaling behaviors may be generic properties of a fully-developed random vortex tangle. These findings may excite future research on hidden scaling laws in QT.

\section*{Results}
\subsection*{Experimental setup and procedures}
Our experimental setup is shown schematically in Fig.~\ref{Fig1}~(a). A 400-$\Omega$ planar resistive heater is installed at the bottom of a vertical flow channel (1.6$\times$1.6$\times$33~cm$^3$) inside a He II bath. The temperature $T$ of the He II in the bath can be controlled by regulating the vapor pressure. When a DC voltage is applied to the heater, a counterflow of the two interpenetrating fluid components of He II establishes in the flow channel~\cite{Landau-book}: the viscous normal-fluid component that consists of thermal quasiparticles in He II (i.e., phonons and rotons) flows away from the heater at a mean velocity given by $v_n$=$q/\rho sT$, where $q$ is the heat flux, and $\rho$ and $s$ are the He II density and specific entropy, respectively; while the inviscid superfluid component (i.e., the condensate) moves in the opposite direction at a velocity $v_s$=$-v_n\rho_n/\rho_s$, where $\rho_n/\rho_s$ is the density ratio of the two fluids.

It has been known that above a small threshold heat flux of the order 10 mW/cm$^2$ \cite{Vinen-1957-PRS-IV}, turbulence appears spontaneously in the superfluid as a random tangle of quantized vortex lines, each carrying a quantized circulation $\kappa\simeq10^{-3}$ cm$^2$/s around its angstrom-sized core~\cite{Donnelly-1991-B}. A mutual friction between the two fluids arises due to scattering of the thermal quasiparticles off the vortices~\cite{Vinen-1957-PRS-III}. Above a heat flux of the order 10$^2$ mW/cm$^2$, the normal fluid can also become turbulent \cite{Marakov-2015-PRB, Gao-2016-PRB, Gao-2017-PRB}, rendering a complex doubly turbulent system~\cite{Gao-2016-JETP,Gao-2017-JLTP,Gao-2018-PRB,Bao-2018-PRB}. Our current research focuses on the low heat flux regime where only the superfluid is turbulent.

To probe the flow, we adopt a particle tracking velocimetry (PTV) technique using solidified D$_2$ particles as tracers. Due to their small sizes (i.e., about 4 $\mu$m in diameter \cite{Mastracci-2018-RSI}), these particles have a small Stokes number in the normal fluid and hence are entrained by the viscous normal-fluid flow \cite{Tang-2020-PRF}. But when they are close to the vortex cores, a Bernoulli pressure due to the superfluid flow induced by the vortex cores can push the particles toward the cores \cite{Donnelly-1991-B}, resulting in the trapping of the particles on the quantized vortex lines. These tracer particles are illuminated by a thin continuous-wave laser sheet and their positions are recorded by a video camera at 90 Hz. We have also installed a pair of second-sound transducers for measuring the vortex-line density using a standard second-sound attenuation method \cite{Sherlock-1970-RSI}. More details can be found in the Method section.

As we reported in Ref.~\cite{Mastracci-2018-PRF,Mastracci-2019-PRF,Mastracci-2019-PRF-2}, two distinct groups of particles can be observed at $q$ below about 10$^2$ W/cm$^2$ (see Fig.~\ref{Fig1}~(b)). The G1 group includes particles entrapped on vortices, resulting in irregular trajectories. The G2 group includes untrapped particles entrained by the up-moving laminar normal fluid, resulting in relatively straight trajectories. The streamwise particle velocity distribution based on the analysis of all trajectories exhibits two nearly separated peaks (see Fig.~\ref{Fig1}~(c)), which allows us to distinguish these two groups of particles for separately analyzing their motion \cite{Mastracci-2018-PRF}. The mean velocity of the G2 particles equals the expected normal-fluid velocity. The G1 particles are carried by the vortex tangle which drifts at $v_s$ towards the heater at small $q$~\cite{Wang-1987-PRB}, but in general the G1 particles may also slide along the vortices due to the viscous drag from the normal fluid.

Nevertheless, at $q$ less than a few tens of mW/cm$^2$, we find that the mean velocity of the G1 particles is about $v_s$~\cite{Mastracci-2018-PRF}, in agreement with the observations of Paoletti \emph{et al.}~\cite{Paoletti-2008-PRL,Paoletti-2008-JPSJ}. This observation suggests that the viscous drag effect on the trapped G1 particles should be mild in the low heat-flux regime. Furthermore, a recent theoretical work suggests that micron-sized tracers trapped on quantized vortex lines in He II are indeed immobilized along the lines due to an effective friction originated from the breakdown of the vortex coherence \cite{Skoblin-2020-JLTP}. Therefore, in the low heat flux regime, it is feasible to explore the genuine vortex-tangle dynamics by tracking the motion of the trapped G1 particles. Specifically, we focus on studying the apparent diffusion of the G1 particles in the horizontal direction so as to keep the viscous drag influence minimal.

\subsection*{Vortex diffusion statistics}
\begin{figure}
\centering
\includegraphics[width=1\linewidth]{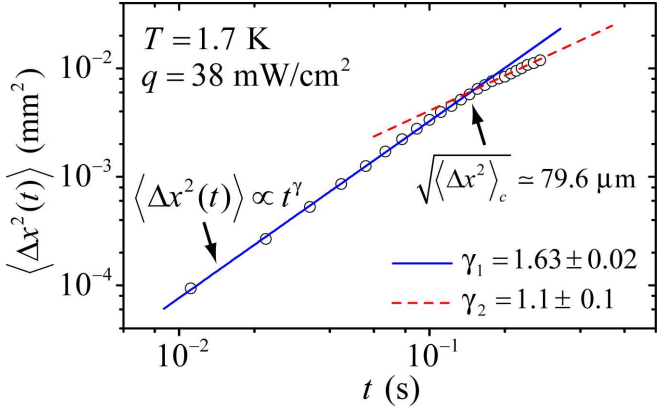}
\caption{Representative data showing the horizontal mean squared displacement $\langle\Delta x^2(t)\rangle$ of the G1 particles as a function of the diffusion time $t$. The solid and the dashed lines are power-law fits to the data.}
\label{Fig2}
\end{figure}

For the data obtained at each temperature and heat flux, we first calculate the horizontal mean squared displacement of the G1 particles $\langle\Delta x^2(t)\rangle$=$\langle[x(t)-x(0)]^2\rangle$, where the diffusion time $t$ starts from the moment when a particle is first observed along its trajectory, and the angle brackets denote an ensemble average over at least $10^3$ trajectories. In general, a power-law scaling $\langle\Delta x^2(t)\rangle\propto t^\gamma$ is expected, where the exponent $\gamma$ is often used to identify different types of diffusions, i.e., normal diffusion ($\gamma$=1), superdiffusion ($\gamma$$>$1), and subdiffusion ($\gamma$$<$1)~\cite{Ben-2000-book}. Fig.~\ref{Fig2} shows a representative result obtained at $T$=1.7 K and $q$=38 W/cm$^2$. The data exhibit two power-law scaling regimes: a superdiffusion regime with $\gamma_1$$\simeq$1.63 at small RMSD (i.e., $\sqrt{\langle\Delta x^2}$) and a nearly normal diffusion regime with $\gamma_2$$\simeq$1.1 at larger RMSD. These two regimes intersect at $\sqrt{\langle\Delta x^2\rangle_c}\simeq79.6$ $\mu$m. Due to their irregular trajectories, the G1 particles seldom stay in the thin laser plane for long time. Therefore, we have relatively few long trajectories to study the G1 particle diffusion at large $t$, which limits the range of the observed $\gamma_2$-scaling regime. We would also like to comment that at sufficiently small diffusion times, the vortex segments are expected to move ballistically at the local superfluid velocity \cite{Donnelly-1991-B}, which should lead to a $t^2$ scaling of $\langle\Delta x^2\rangle$. However, this distinct regime likely would occur only below a few milliseconds for the vortex-line density examined in our experiments, which is beyond the resolution of typical He II PTV measurements.
	
\begin{figure}[h]
\includegraphics[width=1\linewidth]{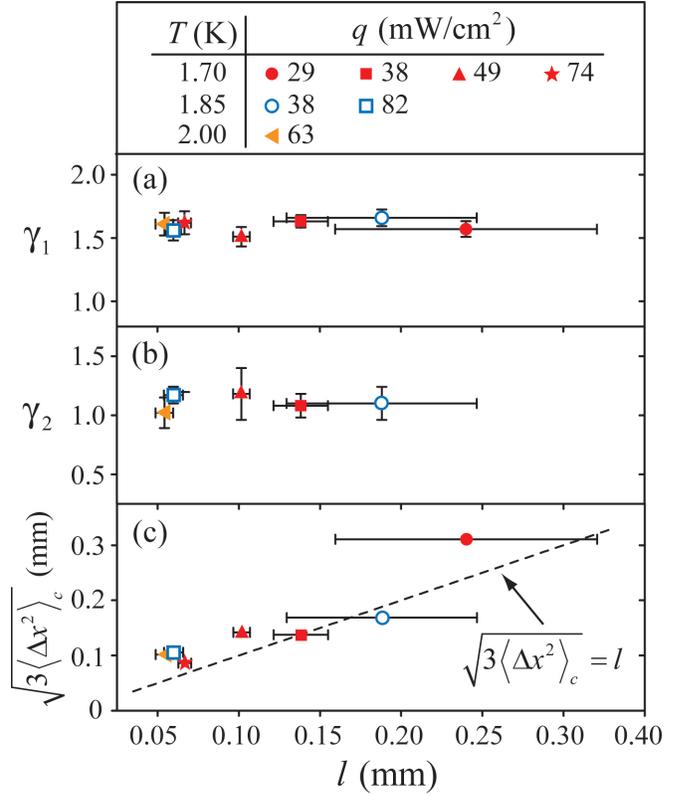}
\caption{(a) and (b) show, respectively, the obtained scaling exponents $\gamma_1$ and $\gamma_2$ for the two diffusion regimes of the G1 particles. (c) A comparison of the transition RMSD with the mean inter-vortex distance $\ell$. The vertical error bars represent the uncertainties in the power-law fits as shown in Fig.~\ref{Fig2}, and the horizontal error bars denote the standard deviation of the measured $\ell$. At small vortex-line density $L$ (i.e., large $\ell$), the increased measurement uncertainty leads to large horizontal error bars.}
\label{Fig3}
\end{figure}

Our analysis of the data sets obtained at other heat fluxes and temperatures also show similar two power-law scaling regimes. The derived $\gamma_1$, $\gamma_2$ and $\sqrt{\langle\Delta x^2\rangle_c}$ are collected in Fig.~\ref{Fig3}. Surprisingly, in the explored temperature range of 1.7 K to 2.0 K where $\rho_n/\rho_s$ varies from 0.3 to 1.24 \cite{Donnelly-1998-JPCRD}, the $\gamma_1$ value is always around $1.6-1.7$ while $\gamma_2$ is close to unity, regardless of the applied heat fluxes. This suggests that the observed diffusion scalings could be generic properties of an evolving random vortex tangle. We have also examined the transition RMSD that separates the two diffusion regimes and find that $\sqrt{\langle\Delta x^2\rangle_c}$ increases with decreasing the vortex-line density $L$. Considering the fact that the diffusion occurs in three-dimensional (3D) space, we compare $\sqrt{3\langle\Delta x^2\rangle_c}$ with the measured inter-vortex distance $\ell$ in Fig.~\ref{Fig3}~(c). These two quantities agree reasonably, suggesting that the transition occurs when the RMSD of the vortices is greater than $\ell$.

It is worthwhile noting that the spatial spreading of a decaying isolated vortex tangle at $T$=0 K either near a solid surface \cite{Tsubota-2003-PB} or in bulk He II \cite{Rickinson-2019-PRB} has been simulated. The growth of the tangle diameter $d$ as reported in Ref.~\cite{Rickinson-2019-PRB} exhibits a normal diffusion regime at large $t$ and a clear superdiffusion regime at small $t$ with a fitted scaling of about $d^2\propto t^{1.74}$ (note that the authors interpreted this latter regime as the ballistic regime). This similarity is encouraging, although our work focuses on the trajectory statistics of vortices in a steady tangle at finite temperatures. Furthermore, following Ref.~\cite{Rickinson-2019-PRB}, we can use our data in the normal diffusion regime to evaluate the effective diffusion coefficient $\nu'/\kappa$=3$\langle\Delta x^2(t)\rangle/4t$. For the data shown in our Fig.~\ref{Fig2}, we get $\nu'/\kappa\simeq0.3$, which is close to the simulated values \cite{Tsubota-2003-PB,Rickinson-2019-PRB}.

\subsection*{Vortex-displacement distribution}
\begin{figure}[t]
\includegraphics[width=1.0\linewidth]{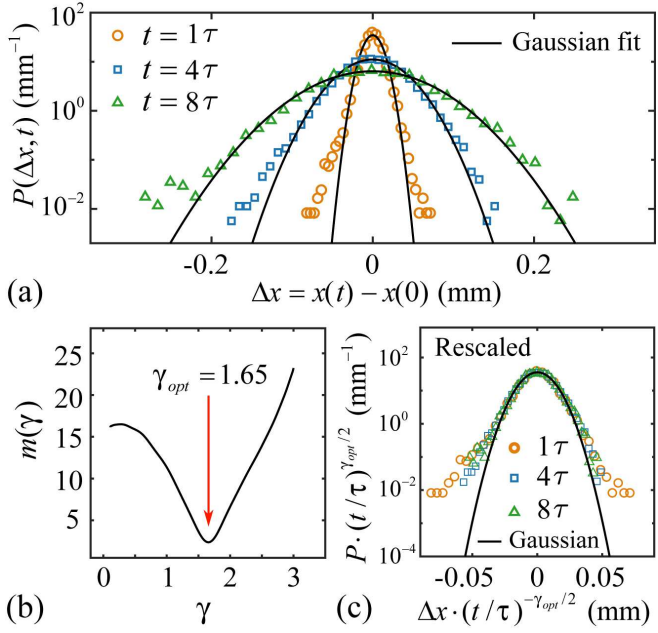}
\caption{(a) Representative particle-displacement distribution function $P(\Delta x,t)$ for the data obtained at 1.7 K and 29 mW/cm$^2$. (b) The measure of the self-similarity $m$ versus the scaling exponent $\gamma$. (c) The rescaled profiles of $P(\Delta x,t)$.}
\label{Fig4}
\end{figure}

Naturally, one would wonder about the cause of the observed vortex-line superdiffusion and why there is a transition to normal diffusion at $\sqrt{3\langle\Delta x^2\rangle_c}\sim\ell$. Indeed, superdiffusion has been observed in a wide range of systems, such as the motion of cold atoms in an optical lattice~\cite{Sagi-2012-PRL}, the chaotic drifting of tracers in rotating flows~\cite{Solomon-1993-PRL}, the cellular transport in biological systems~\cite{Harris-2012-Nature}, and even the search patterns of human hunter-gatherers~\cite{Raichlen-2014-PNAS}. A useful function for characterizing superdiffusion is the distribution function $P(\Delta{x},t)$ of the particle displacement $\Delta{x}$ at time $t$, whose time evolution is often described by a fractional diffusion equation~\cite{Metzler-2000-PR}. A general property of $P(\Delta{x},t)$ is the existence of a self-similar scaling $P(\Delta{x},t)$=$(t/t')^{-\frac{\gamma}{2}}\cdot P(\Delta{x}\cdot(t/t')^{-\frac{\gamma}{2}},t')$, where the scaling exponent $\gamma$ should be identical to the diffusion exponent of the mean squared displacement~\cite{Zaburdaev-2015-RMP}. To test whether this property holds for the apparent diffusion of the G1 particles, we examine the $P(\Delta{x},t)$ profiles at different $t$ for the data taken at 1.7 K and 29 mW/cm$^2$ (see example profiles in Fig.~\ref{Fig4}~(a)). We use the profile at $t=\tau$ as the reference, where $\tau$$\simeq$11 ms is the time step set by the camera frame rate. To determine the optimal exponent $\gamma_{opt}$ that gives the best match among the $P(\Delta{x},t)$ profiles after the rescaling, we minimize the profile difference by calculating the standard $L1$-type variance~\cite{Sagi-2012-PRL}:	
	\begin{equation}
	\resizebox{.88\hsize}{!}{$
		m(\gamma)=\sum\limits_{t=\tau}^{N\tau} \frac{\int\mid (t/\tau)^{\frac{\gamma}{2}}\cdot P(\Delta x\cdot(t/\tau)^{\frac{\gamma}{2}},t)-P(\Delta x,\tau)\mid dx}{\int P(\Delta x,\tau)dx}
		$},
	\label{Eq1}
	\end{equation}	
where the summation goes over all the $P(\Delta{x},t)$ profiles obtained at $t$$\in$$[\tau,N\tau]$, where $N\tau$ is the maximum diffusion time in the $\gamma_1$-scaling regime. Fig.~\ref{Fig4}~(b) shows the calculated variance $m$ as a function of $\gamma$. The minimum $m$ is achieved at $\gamma_{opt}$=1.65, which is indeed close to the diffusion scaling exponent $\gamma_1$=1.57 for the chosen data set. As shown in Fig.~\ref{Fig4}~(c), the rescaled $P(\Delta x,t)$ profiles overlap very well except perhaps in the tail region.

Indeed, for superdifussion systems involving random walkers, another important property of $P(\Delta x,t)$ is its non-Gaussian tails. It has been identified that superdiffusion in those systems is caused by long-distance hops of the walkers ~\cite{Sagi-2012-PRL,Solomon-1993-PRL,Harris-2012-Nature,Raichlen-2014-PNAS,Metzler-2000-PR}, i.e., the so-called L\'{e}vy flights~\cite{Zaburdaev-2015-RMP}. These flights lead to asymptotic power-law tails of the step-displacement distribution $P(\Delta x,\tau)$$\propto$~$|\Delta x|^{-\alpha}$ with $\alpha<3$~\cite{Bouchaud-1990-PR}. After many steps, the resulted $P(\Delta x,t)$ converges to a L\'{e}vy distribution with similar power-law tails. The variance $\langle \Delta x^2\rangle$ for such a heavy-tailed distribution diverges, but a pseudo-variance behavior $\langle \Delta x^2(t)\rangle$$\propto$~$t^{\gamma}$ with $\gamma$=$\frac{2}{\alpha-1}$ can be derived through a scaling argument~\cite{Metzler-2000-PR,Zaburdaev-2015-RMP,Bouchaud-1990-PR}, resulting in an apparent superdiffusion (i.e., $\gamma>1$ when $\alpha<3$). Without such fat tails (i.e., if $\alpha\geq3$), $\langle \Delta x^2\rangle$ would converge, which then leads to a Gaussian distribution of $P(\Delta x,t)$ and hence a normal diffusion of the walkers according to the central limit theorem~\cite{Bouchaud-1990-PR}.

\begin{figure}[t]
\includegraphics[width=1.0\linewidth]{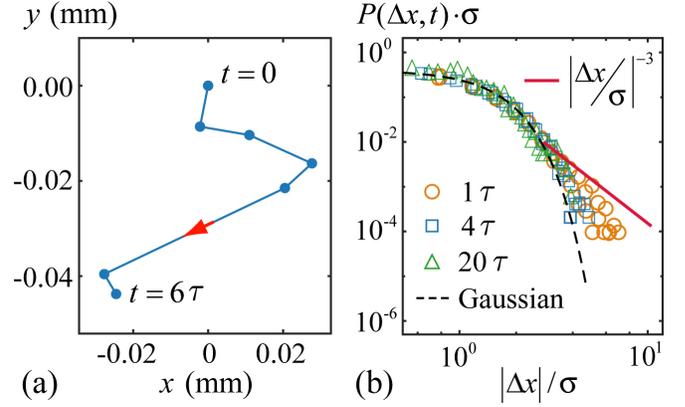}
\caption{(a) An example G1 particle trajectory that exhibits one large step displacement. (b) The tail profiles of $P(\Delta x,t)$, normalized by the standard deviation $\sigma$ of $\Delta x$, at different $t$. The data were taken at 1.7 K and 29 mW/cm$^2$.}
\label{Fig5}
\end{figure}	

Interestingly, the trapped G1 particles do exhibit occasional long-distance hops over the time step $\tau$. An example G1 trajectory that contains an exceptionally large step displacement is shown in Fig.~\ref{Fig5}~(a). The origin of these long-distance hops has been understood as due to the particles carried by vortex segments that are close to locations of vortex reconnections~\cite{Mastracci-2019-PRF}. As revealed by Paoletti \emph{et al.}~\cite{Paoletti-2008-PRL}, vortex reconnections result in local high vortex-velocity occurrences, which lead to non-Gaussian $|v|^{-3}$ tails of the vortex-line velocity distribution. Therefore, when $\tau$ is small, the step-displacement distribution of the vortex lines should acquire similar power-law tails $P(\Delta x,\tau)$$\propto$~$|\Delta x|^{-3}$. However, since the velocities of the reconnecting vortex segments become high only within a short time window centered at the moment of reconnections, over longer time $t$ the total displacement of a vortex segment $\Delta{x}$=$\int_0^t{v_x(t')dt'}$ would rarely exhibit exceptionally large values. Therefore, the tails of the resulted $P(\Delta x,t)$ are suppressed. To see this effect, we show the tails of $P(\Delta x,t)$ at different $t$ in Fig.~\ref{Fig5}~(b) for the data taken at 1.7 K and 29 mW/cm$^2$. Obviously, as $t$ increases from $\tau$ to 20$\tau$, the tail changes from close to $|\Delta x|^{-3}$ to nearly a Gaussian form. This observation is similar in nature to what was reported in Ref.~\cite{La_Mantia-2014-EL}. Therefore, despite the existence of some long-distance hops of the G1 particles at small time steps, their statistical weight is not sufficient to render the observed superdiffusion.

\subsection*{Vortex-velocity correlation}
Without invoking L\'{e}vy flights, superdiffusion may still emerge if the motion of the particles is not completely random but instead exhibits extended temporal correlations~\cite{Bouchaud-1990-PR,Davison-1989-PRS}. For quantized vortices in a vortex tangle, the chaotic motion of the vortex segments is driven by their self- and mutually induced velocities~\cite{Donnelly-1991-B}. There is no existing knowledge on whether this motion is completely random or indeed has a certain temporal correlation.

\begin{figure}[t]
\includegraphics[width=1\linewidth]{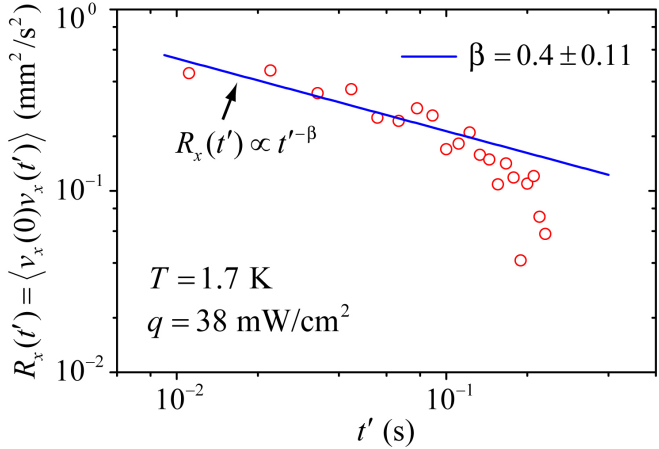}
\caption{The calculated horizontal-velocity temporal correlation function $R_x(t')$ for the data taken at 1.7 K and 38 mW/cm$^2$. The solid line represents a power-law fit.}
\label{Fig6}
\end{figure}

Mathematically, the mean squared displacement $\langle\Delta x^2(t)\rangle$ of a vortex-line segment can be evaluated based on its velocity $v_x(t)$ as~\cite{Mazzitelli-2004-NJP}:
\begin{equation}
\langle\Delta x^2(t)\rangle=2\int_0^tdt_0\int_0^{t-t_0}dt'\langle v_x(t_0)v_x(t_0+t')\rangle,
\label{Eq2}
\end{equation}
where the horizontal-velocity temporal correlation function $R_x(t',t_0)$=$\langle v_x(t_0)v_x(t_0+t')\rangle$ for statistically steady and homogeneous systems would only depend on the lapse time $t'$, i.e., $R_x(t')$=$\langle v_x(0)v_x(t')\rangle$. In this situation, if a power-law scaling $R_x(t')$$\propto$$(t')^{-\beta}$ exists, one can easily derive from equation (\ref{Eq2}) that the mean squared displacement will scale as $\langle\Delta x^2(t)\rangle$$\propto$~$t^{2-\beta}$. On the other hand, if $R_x(t')$ drops rapidly with $t'$, a normal diffusion can be obtained. In Fig.~\ref{Fig6}, we show the calculated $R_x(t')$ for the representative data set included in Fig.~\ref{Fig2}. At small lapse time $t'$, the data do exhibit a power-law scaling with $\beta$ of about 0.4. This scaling exponent leads to $\langle\Delta x^2(t)\rangle$$\propto$~$t^{1.6}$, which agrees nicely with the observed superdifussion. Furthermore, $R_x(t')$ drops sharply beyond a transition time that coincides with the transition to the normal diffusion as seen in Fig.~\ref{Fig2}, which naturally explains this transition. Similar $R_x(t')$ scaling behaviors are also observed for other data sets. These observations provide a direct evidence showing the existence of a possible generic power-law scaling of the vortex-velocity temporal correlation at scales less than $\ell$ for an random vortex tangle.
	
\section*{Discussion}
The analyses that we have presented support the following simply physical picture. The trapped G1 particles move with the quantized vortices in the tangle whose velocities exhibit a power-law temporal correlation. This correlation leads to an apparent superdiffusion of the vortex lines in 3D space. But when their RMSD becomes greater than the mean inter-vortex distance $\ell$, the vortices are expected to move across each other and hence would undergo reconnections. Following the reconnections, the resulted vortex lines move apart towards directions that are distinct from their original directions \cite{Bewley-2008-PNAS,Fonda-2019-PNAS}, a process that effectively randomizes the motion of the vortices. This randomization then leads to a sharp drop of the vortex-velocity temporal correlation and hence results in the normal diffusion of the vortices at large length and time scales.

Note that the spatial velocity correlation functions of vortices in atomic condensates have been simulated \cite{White-2010-PRL,Cidrim-2017-PRA}, where the authors reported a rapid decay of the correlation over a length scale comparable to $\ell$. But to test the physical picture we have outlined, numerical simulations similar to Ref.~\cite{Yui-2020-PRL} need to be conducted so that the temporal correlation of the vortex velocity in steady counterflow turbulence can be examined. Indeed, our communication with the authors of Ref.~\cite{Yui-2020-PRL} has returned encouraging news that their recent simulation does reproduce the power-law scaling of the vortex-velocity correlation as depicted in Fig.~\ref{Fig6}. These authors also notice that the derived diffusion exponent $\gamma_1$ is nearly temperature independent, thereby supporting our observation about the generic nature of this diffusion scaling.

In summary, our work demonstrates that examining the velocity correlations and trajectory statistics of individual vortices in a vortex tangle could uncover hidden scaling properties of QT. Along the lines, many intriguing questions may be raised. For instance, what is the mechanism underlying the observed power-law scaling of the vortex-velocity temporal correlation for a random tangle? Does this scaling also hold for a vortex tangle with large-scale polarizations? How does superfluid parcels undergo apparent diffusion and dispersion in QT? We hope that these questions will stimulate more future researches on vortex and superfluid dynamics.

\matmethods{
\subsection*{Particle tracking velocimetry}
We use solidified deuterium (D$_2$) particles as tracers in He II. These tracer particles are produced by slowly injecting a mixture of 5\% D$_2$ gas and 95\% $^4$He gas directly into the He II bath via a gas injection system similar to what Fonda \emph{et al.} reported \cite{Fonda-2016-RSI}. Upon the injection, the D$_2$ gas forms small ice particles with a mean diameter of about 4~$\mu$m, as determined from their settling velocity in quiescent He II~\cite{Mastracci-2018-RSI}. Following the particle injection, we then turn on the heater and wait for 10 to 20 s for a steady counterflow to establish in the flow channel. A continuous-wave laser sheet (thickness: 200~$\mu$m, height: 9 mm) passes through the geometric center of the channel to illuminate the particles. The positions of the particles in the illuminated plane are captured by a video camera at 90 frames per second. At a give temperature and heat flux, we took a sequence of 720 images and would typically repeat this data acquisition three times to obtain enough particle trajectories for statistical analyses. A modified feature-point tracking routine \cite{Sbalzarini-2005-JSB} is adopted to extract the trajectories of the tracer particles from the sequence of images. The velocity of a particle can be determined by dividing its displacement from one frame to the next by the frame separation time.

\subsection*{Separation data analysis scheme}
To determine whether a tracer particle belongs to the G1 group (i.e., particles that are trapped on vortices) or the G2 group (i.e., untrapped particles that are entrained by the normal fluid), a separation data analysis scheme is adopted \cite{Mastracci-2018-PRF}. As shown in Fig.~\ref{Fig1} (c), the vertical-velocity distribution based on the analysis of all particle trajectories exhibits two nearly separated peaks at low heat fluxes. Through Gaussian fits to these two peaks, we can determine their respective mean velocities (i.e., $\bar{v}_1$ and $\bar{v}_2$) and the corresponding standard deviations (i.e., $\sigma_1$ and $\sigma_2$). Then, for a particle with a vertical velocity $v_y<\bar{v}_2-a_2\sigma_2$, it is categorized as a G1 particle. Otherwise, if $v_y>\bar{v}_1+a_1\sigma_1$, the particle is treated as a G2 particle. Depending on how far the G1 and the G2 peaks are separated, the coefficients $a_1$ and $a_2$ are adjusted in the range of 2 to 6 to better distinguish the two groups. Most of the particle trajectories can be identified as either the G1 type or the G2 type. At relatively large heat fluxes, some trajectories may appear to be partly the G1 type and partly the G2 type. This is due to the particles originally moving with the normal fluid later getting trapped by vortex lines (i.e., G2 to G1) or the trapped particles getting released during vortex reconnections (i.e., G1 to G2). In the current work, we focus on analyzing the whole and partial trajectories that are identified as the G1 type.

\subsection*{Second-sound attention}		
We measure the volume-averaged vortex-line density $L$ in the flow channel using the standard second-sound attenuation method \cite{Sherlock-1970-RSI}. Due to its two-fluid nature, He II supports two distinct sound modes: an ordinary pressure-density wave (i.e., the first sound) where both fluids move in phase, and a temperature-entropy wave (i.e., the second sound) where the two fluids move out of phase. The second-sound waves can be generated and picked up by oscillating superleak transducers \cite{Sherlock-1970-RSI}. These transducers are essentially parallel plate capacitors with one fixed plate and one flexible plate made of a thin porous membrane coated with an evaporated gold layer. By applying an alternating current to one transducer as shown in Fig.~\ref{Fig1} (a), a standing second-sound wave across the channel can be established, whose amplitude can be measured by the other transducer installed on the opposite channel wall. In the presence of quantized vortices, the amplitude of the second-sound wave is attenuated, and the degree of this attenuation can be used to calculate the vortex-line density $L$ \cite{Mastracci-2018-RSI}. Table~\ref{Tab1} lists our measurement results under various temperatures and heat fluxes.
\begin{table}
		\centering
		\caption{Measured vortex-line density $L$}
		\begin{center}
			\setlength{\tabcolsep}{6mm}{
				\begin{tabular}{cccc}		
					\hline\hline
					\specialrule{0em}{2pt}{2pt}
					T(K)& q (mW$/$cm$^2$) & $L$(cm$^{-2}$) \\
					\specialrule{0em}{2pt}{2pt}
					\hline				
					\specialrule{0em}{2pt}{2pt}
				    1.70  &74 & $(22.5 \pm 1.0) \times 10^3$ \\
                          &49 & $(9.7 \pm 1.0) \times 10^3$  \\
					      &38 & $(5.5 \pm 1.3) \times 10^3$  \\
                          &29 & $(1.7 \pm 1.1) \times 10^3$  \\
					\hline				
					\specialrule{0em}{2pt}{2pt}
					1.85  &82 & $(28.2\pm 3.2)\times10^3$\\
					      &38 & $(2.8 \pm 1.7)\times 10^3$\\
					\hline
					\specialrule{0em}{2pt}{2pt}
					2.00  &63 & $(34.9\pm 6.1)\times 10^3$\\
					\hline\hline
			\end{tabular}}
		\end{center}
\label{Tab1}	
\end{table}

\subsection*{Data Availability}
The analysis results together with the flow visualization data and the second-sound data can be obtained from the corresponding author upon request.
}
\showmatmethods{} 

\acknow{The authors would like to acknowledge the valuable discussions with W. F. Vinen and D. Kivotides. The authors also thank S. Yui, H. Kobayashi, and M. Tsubota for communicating their recent simulation results. This work is supported by the National Science Foundation (NSF) under Grant No. DMR-1807291 and the U.S. Department of Energy under Grant No. DE-SC0020113. The experiment was conducted at the National High Magnetic Field Laboratory at Florida State University, which is supported through the NSF Cooperative Agreement No. DMR-1644779 and the state of Florida.}
\showacknow{} 
	
\bibliography{QT-diffusion-NC}
	
\end{document}